\documentclass[prb,preprint,showpacs,preprintnumbers,amsmath,amssymb]{revtex4}
\usepackage{graphicx}
\usepackage{epsfig}
\usepackage{dcolumn}
\usepackage{bm}
\begin{document}
\title{Flux jumps, Second Magnetization Peak anomaly and the Peak Effect phenomenon in single crystals of $YNi_2B_2C$ and $LuNi_2B_2C$}
\author{D. Jaiswal-Nagar\footnote{Email: shikha@tifr.res.in}, A. D. Thakur, S. Ramakrishnan, A.K. Grover\footnote{Email: grover@tifr.res.in}}
\address{Department of Condensed Matter Physics and Materials Science, Tata Institute of Fundamental Research, Homi Bhabha Road, Colaba, Mumbai-400005, India}
\author{D. Pal}
\address{Department of Physics, Indian Institute of Technology Guwahati, Guwahati, 781039, India}
\author{H. Takeya}
\address{National Institute for Materials Science, Sengen 1-2-1, Tsukuba, Ibaraki 305-0047, Japan}
\begin{abstract}
We present magnetization measurements in single crystals of the tetragonal $YNi_2B_2C$ compound, which exhibit the phenomenon of peak effect as well as the second magnetization peak anomaly for H $>$ 0.5~T (H $||$ c). At the lower field  (50~mT $<$ H $<$ 200~mT), we have observed  the presence of flux jumps, which seem to relate to a structural change in the local symmetry of the flux line lattice (a first order re-orientation transition across a local field in some parts of the sample, in the range of 100~mT to 150~mT). These flux jumps are also observed in a single crystal of $LuNi_2B_2C$ for H $||$ c in the field region from 2 mT to 25 mT, which are compatible with the occurrence of a re-orientation transition at a lower field in a cleaner crystal of this compound, as compared to those of $YNi_2B_2C$. Vortex phase diagrams drawn for H $||$ c in $LuNi_2B_2C$ and $YNi_2B_2C$ show that the ordered elastic glass phase spans a larger part of (H, T) space in the former as compared to latter, thereby, reaffirming the difference in the relative purity of the two samples.
\end{abstract}

\maketitle

\section{INTRODUCTION}
\label{sec:INTRO}
\noindent
A well documented and researched issue in the context of vortex phase diagrams \cite{rblatter1,rgiamarchi2,rkhaykovich4,rledussal3} of both low $T_c$ \cite{rsatyajit5,rcheon6,rklein7,rdivakar8} as well as high $T_c$ superconductors \cite{rkhaykovich4,rsoibel9,ravraham10,rdilip11,rdilip12} is the phenomenon of peak effect (PE) \cite{rgiamarchi2}, which is an anomalous increase in the critical current density ($J_c$) prior to reaching the normal-superconductor phase boundary. The PE is widely considered to mark a first order transition from a collective pinned ordered vortex solid to an individually pinned disordered solid \cite{rgiamarchi2,rling13,rpark14,troyanovski}. Another anomalous feature seen deep in the mixed state of weakly pinned samples of low as well as high $T_c$ superconductors  is the second magnetization peak (SMP) anomaly \cite{rkhaykovich4,rdaeumling15,rshampa16,rshampa17}. The SMP anomaly is often related to the pinning induced transition from dislocation free elastic glass (Bragg glass \cite{rledussal3}) to the dislocation mediated multi-domain vortex glass state. In such a circumstance, the onset field of the SMP anomaly is not expected to vary with temperature \cite{rledussal3,rgingras18}.\\
Abrikosov \cite{rabrikosov19} had predicted the flux lines to be arranged in a regular array, he found the periodic array to be a square, but the difference in energy between the square and a triangular array is only 2 $\%$. Numerous reports of change in the symmetry of the VL from a rhombohedral towards a square symmetry \cite{robst20} observed by small angle neutron scattering (SANS) measurements and Bitter decoration studies \cite{rcanfield21,reskildsen22,rmcpaul23,rvinnikovh224,rvinnikovh125,rlevett26,rdewhurst27} in the quaternary borocarbide compounds \cite{rnag28,rcava29}, have resurged the interest in exploring and understanding  the underlying mechanism governing this process in a wide variety of superconductors, e.g., {\bf Pb}Tl \cite{robst20}, $V_3Si$ \cite{ryethiraj30}, $La_{1-x}Sr_{x}CuO_{4}$ \cite{rgilardi31,rrosenstein32}, Nb \cite{rforgan33}, $YBa_2Cu_3O_7$ \cite{rbrown34}, etc.\\
 The borocarbide superconductors which have a tetragonal crystal structure with c/a $\sim$~3, are  convenient test beds to study the interesting phenomenon of the change in the flux line lattice symmetry and the effects dependent on it, as high quality single crystals of large enough sizes can be grown by different procedures \cite{rcanfield21}. In the non-magnetic members of the borocarbide series, viz., $YNi_2B_2C$ (Y1221) and $LuNi_2B_2C$ (Lu1221), it is observed that for H $||$ [001] at low fields ($\sim$ 0.1 T), the VL symmetry is a distorted triangle (lower field rhombohedral $R_L$) with an apex angle $\beta_1 < 60^0$. With the increase in field, the VL undergoes a sudden (first order) transformation via a $45^0$ re-orientation to higher field rhombohedral $R_H$, with an apex angle $\beta_2$ ($>~60^0$). The $R_H$ symmetry subsequently smoothly proceeds to a square symmetry via a continuous (second order) transition at a field $H_2$ \cite{reskildsen22,rmcpaul23}. In the typical crystals of Y1221, the $H_1$ field (for H $||$ [001]) lies in the range $\sim$ 100-150~mT \cite{rmcpaul23,rlevett26,rdewhurst27}, whereas in the very clean crystals of Lu1221, the same field lies in the lower field interval, 20-50~mT for H $||$ c \cite{rvinnikovh224,rvinnikovh125}. Further studies \cite{eskildsen,eskildsenpra} in crystals of $LuNi_2B_2C$ have revealed that for H $||$ [100], the rhombohedral $R_H$ (apex angle $\beta_2 > 60^0$) undergoes a sudden reorientation transition at $H_{tr} \sim$~300~mT, such that the body diagonal of distorted rhombohedral locks up in [010] direction with apex angle $\sim 82^0$. \\
Kogan {\it et al.} \cite{rkogan35} predicted the occurence of the discontinous (first order like) $R_L \rightarrow R_H$ transition at $\sim$ 20 mT for H $||$ [001] in clean crystals of borocarbides. Subsequent studies by Gammel {\it et al.} \cite{rgammel36} in Co doped Lu1221 crystals revealed that the $R_H$ to square transition for H $||$ [001] shifts from lower to higher fields with the increase in the Co doping, i.e., with the increase in disorder effects. However, the possibilities of relationship(s) between the structural transition(s) in the VL and the spatial order-disorder transitions {\it a la} PE/SMP have not been described. It is of interest to know (i) how the symmetry transformations adjust to the pinning landscape, and (ii) whether the domain volume within which VL remains correlated depends on the underlying symmetry and the disorder effects on it. Magnetization studies by Silhanek {\it et al.} \cite{rsilhane37} on a single crystal of Y1221 with an applied field H $||$ c (i.e., [001]) revealed the presence of a kink in the pinning force density at a field value, which is close to the $H_1$ value in this compound. To our knowledge, no other signature(s) of the VL symmetry transitions have been reported in the magnetization hysteresis measurements, while there have been several reports of the observation of the PE in the samples of Y1221 \cite{reskildsen22,rxu38,rhirata39,rsong40,rjames41} and Lu1221 \cite{reskildsen22}.\\
We report here on the observation of the PE, the SMP anomaly and the flux jumps in the same isothermal magnetization hysteresis scan for H $||$ c in the crystals of Y1221 and Lu1221. The flux jumps interestingly occur in the field regime, where the VL symmetry transition is reported to occur across the respective $H_1$ values for applied field oriented in the c-direction in each of the compounds. We believe that these flux jumps indeed have a correlation with the local symmetry change in the vortex lattice of borocarbide superconductors. To corroborate this possibility, we have traced in several ways, the minor magnetization curves by changing the initial thermomagnetic history of the sample of Y1221. The measurement of the quadrupolar signal, which purports to preferentially fingerprint the inhomogeneity in the magnetization across the sample, also, registers the change in the symmetry of the VL. The loci of the threshold fields at which the SMP anomaly and the PE commence can lead to the demarcation of boundaries across which changes in the spatial correlation of VL occur. We present construction of the vortex phase diagrams for H $||$ c in the crystals of Y1221 and Lu1221. The parametric region in (H, T) space over which elastic glass state exists, seems to be influenced by the purity of the crystal. In Y1221 crystal, elastic glass state spans over a smaller region, indicating stronger pinning effect in this sample as compared to that in the crystal of Lu1221.

\section{EXPERIMENTAL}

Magnetization measurements have been performed on two single crystals of Y1221, labeled as A and B, and a crystal of Lu1221. The single crystals, A and B, of Y1221 were grown by the travelling solvent floating zone method \cite{rtakeya42}, while the single crystal of Lu1221 was grown by the flux method, using $Ni_2B$ as flux \cite{rcanfield21}. The crystal A of Y1221 and that of Lu1221 are (thin) platelets in shape, with the c-axis perpendicular to the plane of the platelet. The  crystal B of Y1221 is, however, a parallelopiped (of size $\sim$ 3~mm (l) $\times$ 0.7~mm  (b) $\times$ 0.67~mm (t)) in shape, with the a-axis along the largest dimension. Both the crystals of Y1221 have $T_c(0)$ $\approx 15.1~K$, whereas the crystal of Lu1221 has a transition temperature, $T_c$(0) $\approx 16.1~K$. The DC magnetization measurements were performed using (i) a 12~Tesla Vibrating Sample Magnetometer (VSM) (Oxford Instruments, U.K.) and (ii) a 7.5~Tesla SQUID magnetometer (Model MPMS7, Quantum Design Inc., U.S.A.).

\section{RESULTS AND DISCUSSIONS}
\subsection{Magnetization hysteresis measurements in $YNi_2B_2C$}
\subsubsection{Peak effect and second magnetization peak anomaly}

\begin{figure}[!htb]
\begin{center}
\includegraphics[scale=0.4,angle=0] {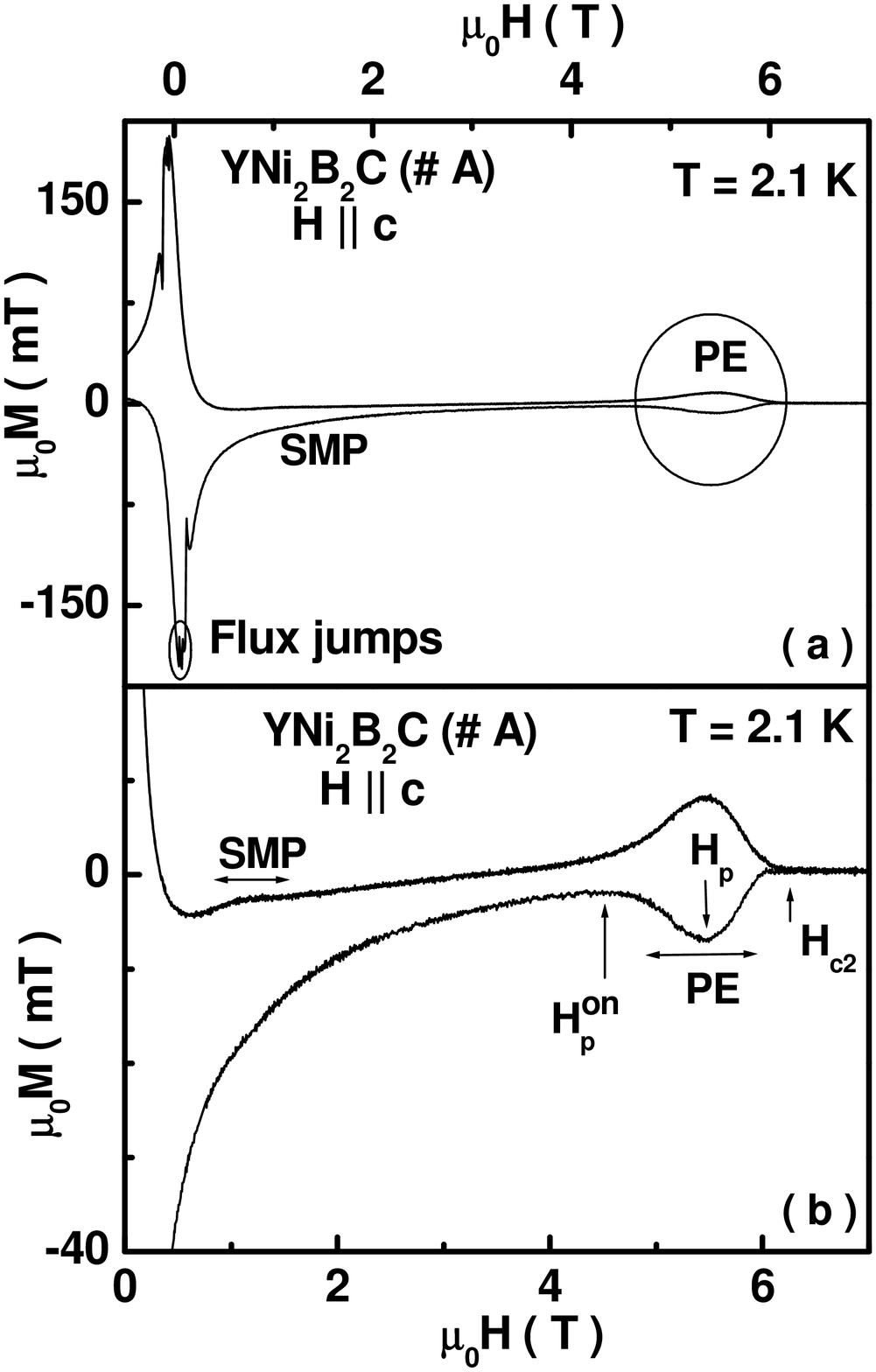}
\caption{A portion of the five quadrant M-H loop at 2.1~K in Y1221 ($\#$ A) with H $||$ c. The presence of a peak effect `bubble' and flux jumps (four in the third quadrant and three in the fifth quadrant) can be noted in the panel (a). Panel (b) shows a blow up of the M-H loop highlighting the second magnetization peak anomaly. The locations of the $H_p^{on}$, $H_p$ and $H_{c2}$ are marked in this panel.}
\label{FIG. 1.}
\end{center}
\end{figure}

The PE in a weakly pinned crystal A of Y1221 is clearly evident (see Fig.~1(a)) by the characteristic `bubble' in a five quadrant M-H loop recorded at 2.1~K using a VSM with a scan rate of 0.35~T/min and with an applied field H $||$ c. Also, evident in Fig.~1 (a) is the presence of flux jumps at fields less than 200~mT, far below the peak effect region. The field values at which the flux jumps occur, vary somewhat from scan to scan, recorded at the same ramp rate of the swept magnetic field. This phenomenon will be described in detail in section B.\\
Fig.~1(b) shows the M-H plot on an expanded scale to emphasize the presence of an anomalous feature designated as the second magnetization peak (SMP) anomaly around a field of 1~T. The large hysteresis in the magnetization at fields less than 0.5~T in Fig.~1(a) gives way  to a much smaller irreversibility at higher fields (upto close to the onset field $H_p^{on}$ of the PE), indicating that the pinning in the sample is weak and an ordered elastic glass phase (i.e., a Bragg Glass (BG) \cite{rledussal3} like state) gets established well before the field regime of the PE. From the hysteresis plot of Fig.~1(a), it may seem that an ordered BG phase extends all the way up to onset field of the PE, but a closer look at the data in Fig.~1(b) (which reveals the presence of a SMP anomaly), suggests that the BG phase could terminate at the onset field of the SMP anomaly above which the multi-domain vortex glass (VG) phase ensues \cite{rkhaykovich4,rledussal3}. The hysteresis width, $\Delta M$(H) ($\propto J_c$(H)), starts to decrease with field once again above the peak field of the SMP anomaly and continues upto the onset field of PE, thereby implying an improvement in the spatial order in this interval. A complete amorphization of the multi-domain elastic VL commences only at $H_p^{on}$.

\begin{figure}[!htb]
\begin{center}
\includegraphics[scale=0.4,angle=270] {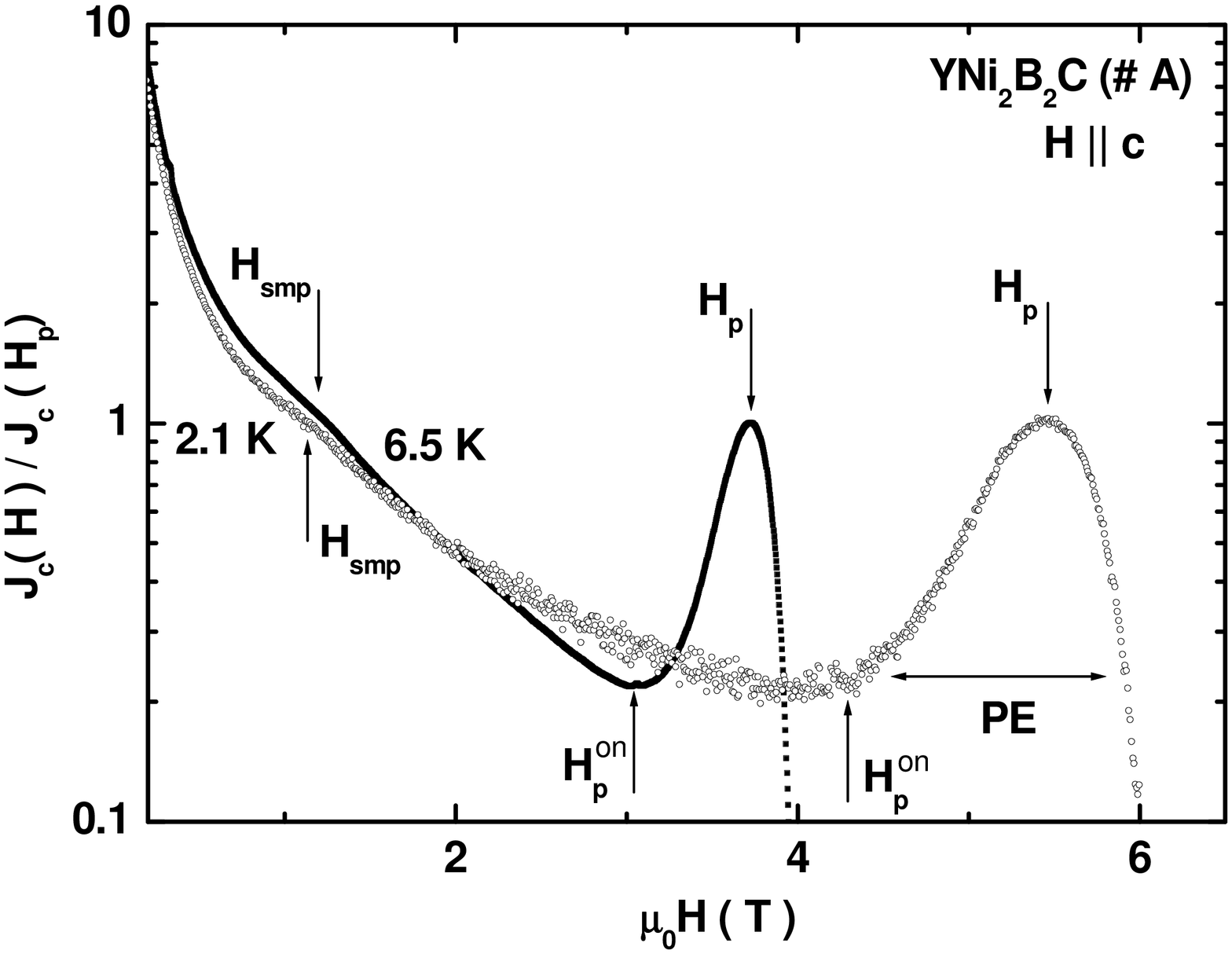}
\caption{Log-linear plots of normalized critical current density, $J_c$(H)/$J_c$($H_p$) ( =  $\Delta$M(H)/$\Delta$M($H_p$)) vs. field ($\mu_0$H) in Y1221 ($\#$ A) at two temperatures (2.1~K and 6.5~K) for H $||$ c. The locations of the maximum position of the SMP anomaly ($H_{smp}$) and the onset ($H_p^{on}$) and maximum ($H_p$) positions of the PE are marked.}
\label{FIG. 2.}
\end{center}
\end{figure}

Fig.~2 shows a (normalized) $J_c$ versus H plot corresponding to the data of Fig. 1(a). Also, shown for comparison, is a (normalized) $J_c$(H) plot (H $||$ c), at a higher temperature ( T = 6.5~K ). The peak fields of the PE ($H_p$) and the SMP ($H_{smp}$) stand marked for both the temperatures. $H_p$ decreases  considerably with an increase in temperature (from 5.4~T at 2.1~K to 3.7~T at 6.5~K), but $H_{smp}$ remains near 1.1~T at both the temperatures. This leads to an inference that the $H_{smp}$(T) line in Y1221 ($\#$ A) would exhibit a very weak temperature dependence, reminiscent of the behavior of the SMP anomaly in another weakly pinned low $T_c$ superconductor, viz., $Ca_3Rh_4Sn_{13}$ ($T_c(0) \approx$ 8.2~K) \cite{rshampa16,rshampa17}.

\begin{figure}[!htb]
\begin{center}
\includegraphics[scale=0.4,angle=270] {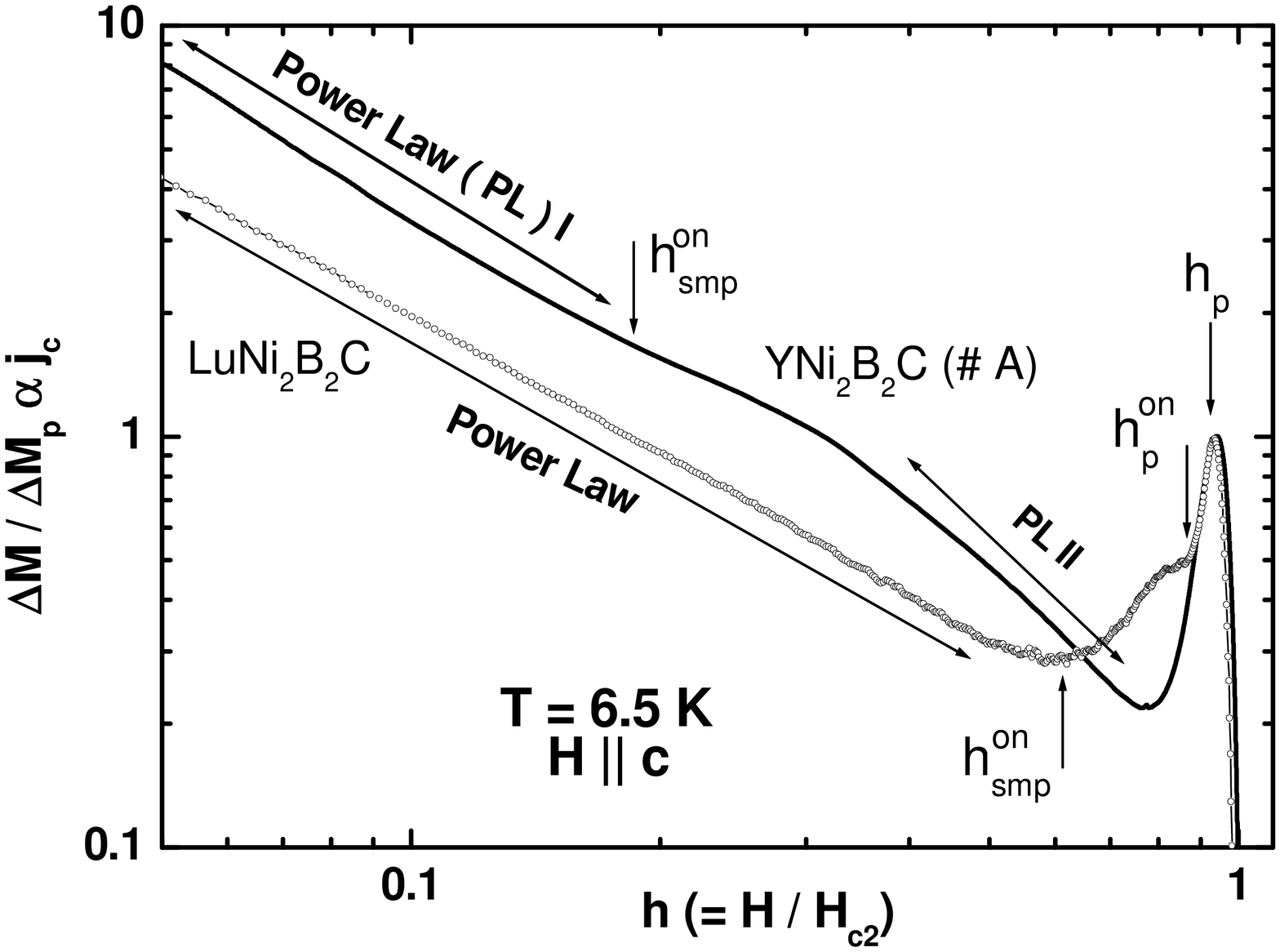}
\caption{Log-log plot of normalized critical current density $\Delta$M/$\Delta$$M_p$($\propto j_c$) vs. normalized field (h=H/$H_{c2}$) in Y1221 ($\#$ A) and Lu1221 at T = 6.5~K for H $||$ c. The peak effect as well as the second magnetization peak anomaly are evident in the data of both the crystals. While only one power law region is evident for the case of Lu1221, two power law regions, viz., PL-I and PL-II can be seen for Y1221.}
\label{FIG. 3.}
\end{center}
\end{figure}

Fig.~3 shows a log-log plot of the current density (normalized to its peak value across the PE, where the correlation volume of VL is expected to reach a  minimum) with field (normalized to the upper critical field $H_{c2}$) in  Y1221 ($\#$ A) with H $||$ c and at 6.5~K. Also, shown for comparison in this figure, is a plot in Lu1221, at (nearly) the same temperature with H $||$ c. Both compounds exhibit SMP like anomaly distinct from the PE feature, but with one little difference. While for Y1221, the onset field of SMP anomaly lying deeper inside the mixed state is well separated from the PE feature and does not display temperature variation, for Lu1221, the SMP anomaly lies at the edge of the PE and varies with temperature as the PE. \\
The field regime where $j_c$ varies with field in a power law manner, i.e., $j_c \propto h^{-n}$, is often  demarcated \cite{rsatyajit5,rshampa16} as the collective pinned elastic regime. In Lu1221, the power law behavior can be seen (cf. Fig.~3) to extend upto h (= H/$H_{c2}$) $\sim$ 0.6, after which the BG phase probably gets broken up into multi-domain VG phase, which starts to amorphize at the onset of PE (h $\sim$ 0.85). On the other hand, for Y1221, one can mark out two regions for the power law behavior. The first power law region (PL I) extends only upto h $\sim$ 0.18 and the second region (PL II) surfaces between h $\sim$ 0.4 and 0.75, where the dislocations injected in the interval, h $\sim$ 0.18 - 0.4, could partially heal. \\
The ratio of $J_c$ at the peak position of the PE to that at the onset of the PE is about five for Y1221, while the same ratio is about two for Lu1221. This implies that the correlation volume in Y1221 shrinks to about 1/25 of its value at the onset of PE ($J_c$ $\propto$ 1/$\sqrt{V_c}$), while for Lu1221, $V_c$ at $H_p$ has shrunk only to 1/4 of its value at $H_p^{on}$. This is plausible, since for Lu1221, the process of reduction in $V_c$ starts at $H^{on}_{smp}$ and it continues till the arrival of $H_p^{on}$. On the other hand, for Y1221, the process of reduction in $V_c$ starts at the onset of PE. Prior to it, the VL in this case has a possibility to heal between $H_{smp}$ and $H_p^{on}$. In fact, the ratio of $J_c$ at $H_p$ to its value at $H_{smp}^{on}$ (h $\sim$ 0.6) is also about five in Lu1221. The premise that the SMP anomaly is disorder induced could imply that its onset at h $\sim$ 0.18 in the given Y1221 crystal ($\#$ A) as compared to the onset of SMP at h $\sim$ 0.6 in Lu1221 crystal, reveals the relative purity (levels of effective disorder) of the samples of these two compounds.

\subsubsection{Pinning force density}

Fig.~4 shows a plot of the pinning force density ($\propto J_c \times$ H) with field on a linear-log scale in Y1221; the data correspond to the magnetization hysteresis measurements shown in Fig.~1(a). Three peaks, marked as $(F_p)^{max}$, SMP and PE, can be clearly distinguished. An increase in the pinning force upto $(F_p)^{max}$ is a representation of the increase in the rigidity of the vortex lattice as a result of the interactions between the flux lines. A peak at $(F_p)^{max}$ probably implies that the rigidity of the vortex  lattice has attained a limiting value dictated by the interaction effects. The second hump corresponds to the SMP anomaly and the third peak corresponds to the quintessential PE \cite{rshampa17}. Apart from the three maxima, one can also mark out the positions of three flux jumps in this figure. These jumps are observed to lie close to, though a little lower, the field value corresponding to the $(F_p)^{max}$. This suggests that flux jumps are observed in the field regime, where the lattice rigidity is effective and VL is well formed.

\begin{figure}[!htb]
\begin{center}
\includegraphics[scale=0.4,angle=270] {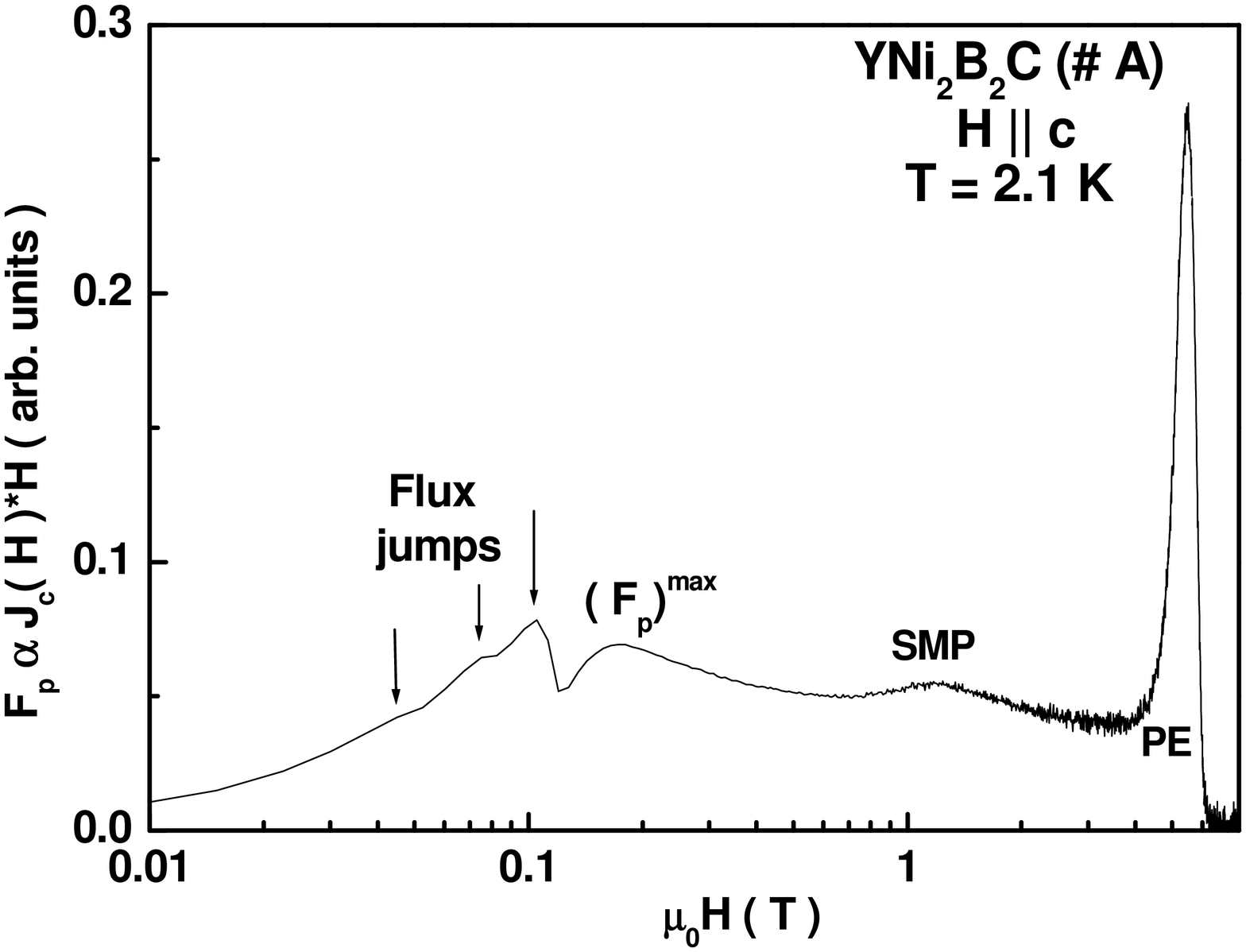}
\caption{Plot of the pinning force density ($J_c$ $\times$H) vs. field in Y1221 ($\#$ A) with H $||$ c. The regimes of $(F_p)$$^{max}$, SMP and PE have been indicated, and the positions of the flux jumps have been marked by arrows.}
\label{FIG. 4.}
\end{center}
\end{figure}

\subsection{Flux jumps in $YNi_2B_2C$}
\subsubsection{Zero field cooled measurements in crystals of Y1221}

Panel (a) of Fig.~5 shows a five quadrant M-H loop recorded at 2.1~K in a VSM at a sweep rate of 0.25~mT/sec in Y1221 ($\#$ A) with H $||$ c. One can clearly discern the presence of the multiple flux jumps in selective quadrants. The following observations are noteworthy:
\begin{figure}[!htb]
\begin{center}
\includegraphics[scale=0.5,angle=0] {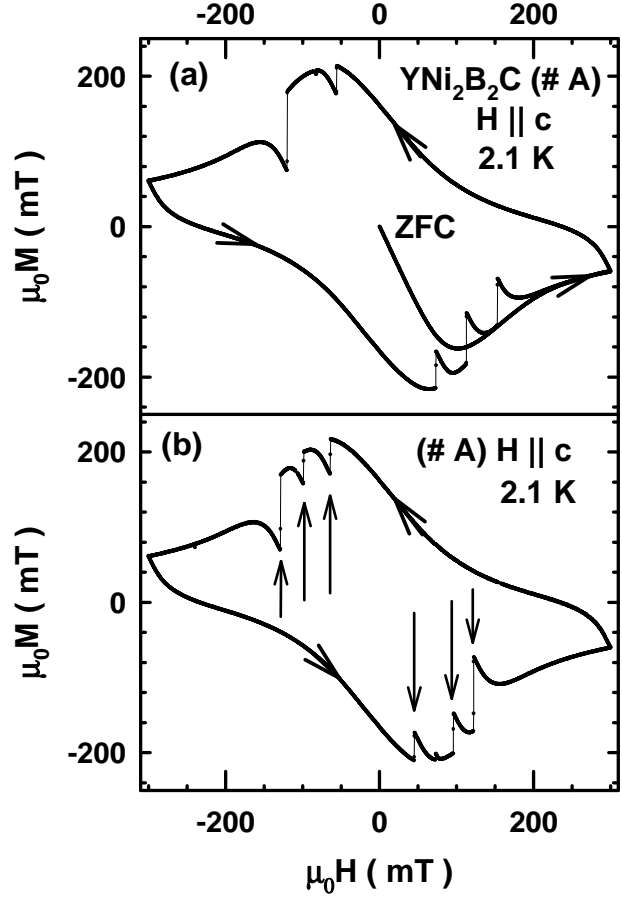}
\caption{Panel (a) shows a five quadrant M-H loop in Y1221 ($\#$ A) at T = 2.1~K with H $||$ c using a VSM with a scan rate of 0.25~mT/sec. Presence of two flux jumps in the third quadrant and three flux jumps in the fifth quadrant can be clearly noted. Panel (b) shows the M-H curve obtained after cycling the field several times between $\pm$$H_{max}$ (= $\pm$ 300~mT). Note that the number of flux jumps in third and fifth quadrant is now same.}
\label{FIG. 5.}
\end{center}
\end{figure}

(i) Flux jumps are absent in the ZFC run (0 $\rightarrow H_{max}$), in the second quadrant ($H_{max}\rightarrow 0$ ) and in the fourth quadrant (-$H_{max} \rightarrow 0$). They occur only in third (0 $\rightarrow -H_{max}$) and fifth (0 to $H_{max}$, subsequent to the initial ZFC run) quadrant.\\
(ii) Two flux jumps occur at about -55~mT and -120~mT, in third quadrant and three flux jumps happen at 70~mT, 110~mT and 150~mT in fifth quadrant. \\
On cycling the field repeatedly between $\pm$ 300~mT several times, it was noted that the number of flux jumps in third and fifth quadrants stabilized to three (see panel (b) in Fig.~5), however, the precise field values at which the jumps happen were found to vary each time. In order to overcome an apprehension that these flux jumps could be an artefact of the rapid ramping of the magnetic field and/or the procedure of magnetization measurement in a VSM, M-H loop was also recorded using a SQUID magnetometer, where the superconducting magnet is kept in the persistent mode while ascertaining the magnetization value of the sample. Qualitatively, the same behaviour (data not shown here) as depicted in Fig.~5 was noted. The number of flux jumps in one of the quadrants (namely, third) were, however, observed to increase from three to four. We are inclined to surmise that the number of jumps in the third/fifth quadrant could statistically vary between two to four.

\begin{figure}[!htb]
\begin{center}
\includegraphics[scale=0.4,angle=270] {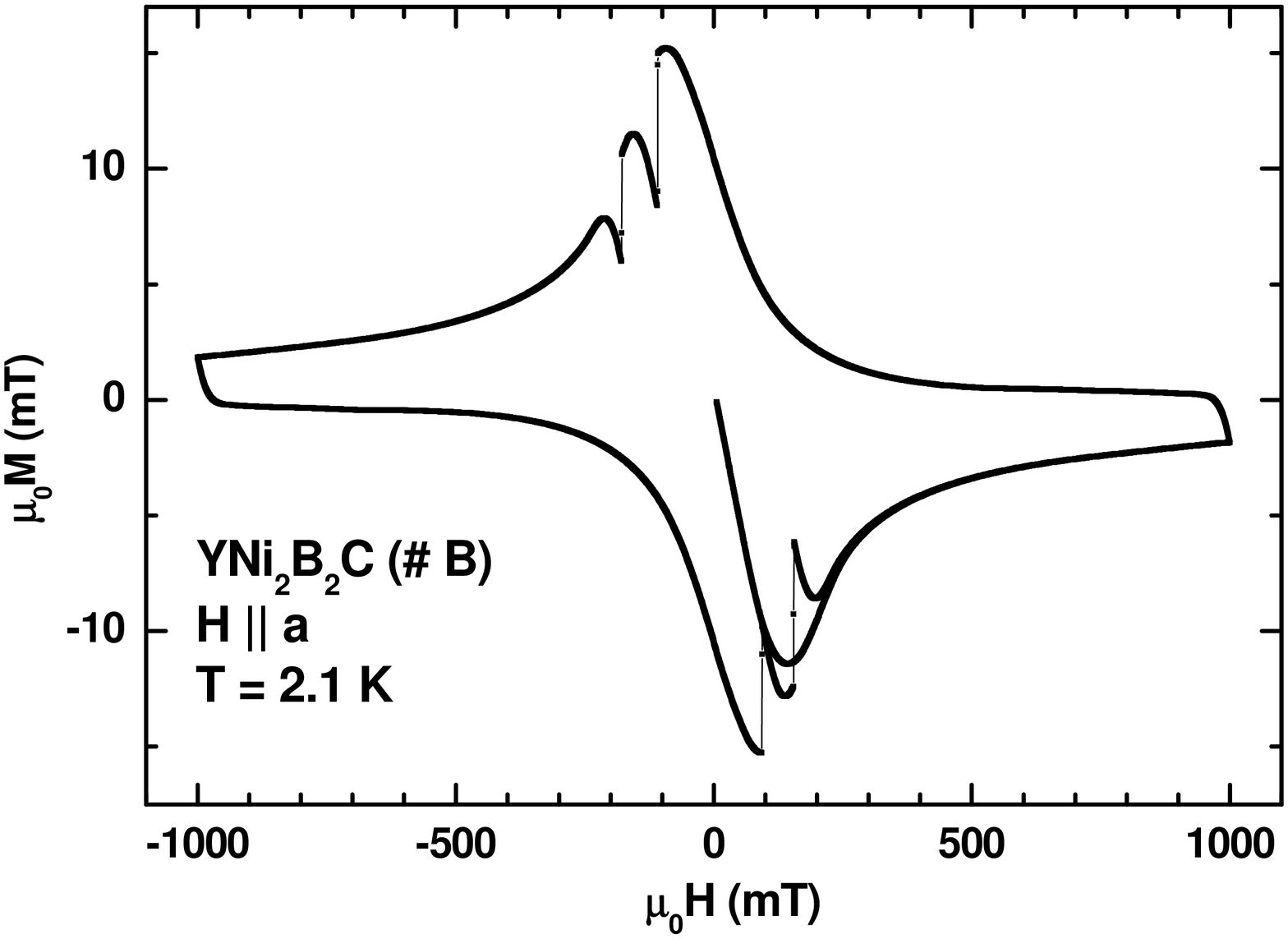}
\caption{M-H loop recorded using VSM (scan rate of field 0.025~T/min) in Y1221 ($\#$ B) at T = 2.1~K with H $||$ a.}
\label{FIG. 6.}
\end{center}
\end{figure}

To establish the notion that the flux jumps do not depend on any specific physical characteristic of a given sample of Y1221, the M-H loops were recorded on another sample of Y1221 (crystal B, which is parallelopiped in shape) with field applied parallel to its longest physical dimension (H $||$ a in this case, see Fig.~6). The presence of flux jumps in the third and fifth quadrants attests to the fact that demagnetization factor of the sample does not influence the manner in which the flux jumps get observed. The observation of flux jumps for H $||$ a orientation in Fig.~6, also, suggests that the mechanism responsible for these jumps probably does not depend on the orientation of the applied field w.r.t. the crystalline axis.

\begin{figure}[!htb]
\begin{center}
\includegraphics[scale=0.4,angle=0] {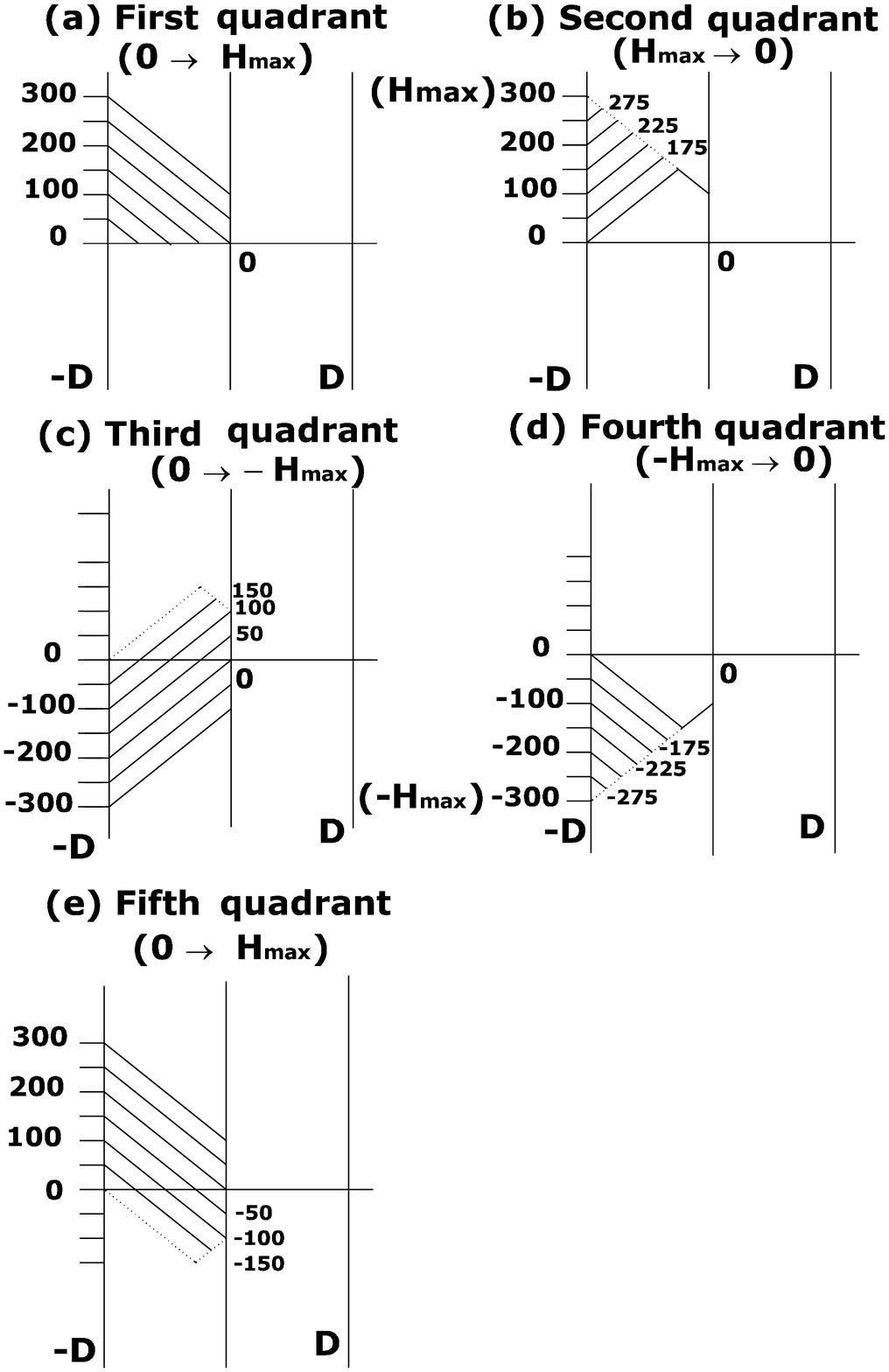}
\caption{Bean's profiles (only one half section (-D to 0) shown) relevant for the five quadrant M-H loop in Fig.~5 (a).}
\label{FIG. 7.}
\end{center}
\end{figure}

It may be pertinent to state here that flux jumps in the third and fifth quadrants have been noted in the hysteresis loops pertaining to the intermediate state of some of the specimen of Type-I superconductors, viz., Pb \cite{rinke}. It is believed that in such a circumstance, the flux jumps are caused by the escape of flux by the fusion of local macroscopic regions of positive and negative magnetization lying in juxtaposition.\\
In order to comprehend the mechanism governing the flux jump process, field profiles relevant to Fig.~5 are drawn, as per a simplified prescription of the Bean's Critical State model \cite{rbean43} for an infinite slab of thickness 2D, with surfaces at x = $\pm$ D, the applied field H being parallel to the surface. We realize that our experimental sample is either platelet shaped or is a parallelopiped, but it is hoped that the Bean's profile drawn in Fig.~7, for zero demagnetization limit and for $J_c$ (H) = constant for the sake of simplicity, would qualitatively turn out to be instructive even when $J_c$ (H) decreases with an increase in H. For brevity, profiles are shown in only one half of the sample. The other half is a mirror image of the drawn profiles. While drawing Bean's field profiles, B-x (local macroscopic field, B, versus distance, x, from the centre of the sample), we have assumed that the full penetration field, $H^*$, is 200~mT. In the prescription of Bean's model, $H^*$ is the limiting field at which the applied field invades the entire sample, after initial zero field cooling the sample (see panel (a) of Fig.~7). The estimate of $\sim$ 200~mT for $H^*$ in Y1221, with H $||$ c, at 2.1~K is based on the magnetization data in Fig.~5 (a). It is the limiting field where the virgin ZFC magnetization curve would merge with the forward leg of the envelope \cite{rchaddah44} loop in the absence of the flux jumps. When $J_c$ (H) decreases with field, such a limiting field can be taken to identify the nominal $H^*$ value.\\
In Fig.~7, we have restricted $\pm$ $H_{max}$ to $\pm$ 300~mT in conformity with the M-H data shown in Fig.~5(a). Panels (c) to (e) of Fig.~7 are relevant for the M-H loop shown in panel (b) of Fig.~5 as well. It may be noted that the field profiles in panels (a), (b) and (d) of Fig.~7 are such that the magnetic field inside the sample remains of the same sign (positive or negative) in these cases. The field profiles in panels (c) and (e) of Fig.~7 allow for the possibility of both positive and negative field values inside the sample. The positive and negative (macroscopic) fields may be identified with the domains (or local regions) comprising vortices and anti-vortices. A narrow region encompassing zero field value would be free of any kind of vortices, and in its neighborhood, the domains of vortices and anti-vortices would lie in juxtaposition. The fact that the flux jumps are observed in the third and the fifth quadrants in conjunction with the Bean's profiles in Fig.~7 could further imply that co-existence of domains of vortices and anti-vortices is necessary for the jumps to occur.\\
We would like to now surmise that mere juxtaposition of the regions of vortices and anti-vortices is not adequate to trigger flux jumps. We conjecture that sudden annihilation of vortices (flux jumps) gets triggered, when a change in the local symmetry of the vortex lattice from low field rhombohedral $R_L$ to higher field rhombohedral $R_H$ (or vice-versa) occurs over a portion of the sample (i.e., in some domains), while the regions of positive and negative fields lie in juxtaposition anywhere else in the sample. Let us call the domains comprising rhombohedral $R_L$ and negative(positive) field as $R_L^{-(+)}$, and, similarly, the domains with rhombohedral $R_H$ and negative(positive) field as $R_H^{-(+)}$. The $R_L^+ (R_L^-) \rightarrow R_H^+ (R_H^-)$ (and vice-versa) transition(s) can be expected to happen in Y1221 crystal as the (macroscopic) field inside the sample crosses the region of 100-150 mT during the ramping of the external field. In addition, the $R_H$ domains can, however, exist in (metastable) supercooled state during the field cooling process even in a field value somewhat less than 100~mT, as has been reported in some of the SANS experiments \cite{rmcpaul23,rlevett26}. Such metastable domains could display the tendency to suddenly transform to the stable domains, during the subsequent field ramping cycle.\\
 On the basis of the above conjecture, let us now reexamine the observations in Fig.~5. During the first ramp up of the field after ZFC (0 mT to 300 mT), the disordered bundles of vortices will enter the sample from the corners and edges and attempt to settle down into regions of $R_L^+$ and $R_H^+$ domains, with the former lying near the centre for H = 300~mT (see Fig.~7 (a)). As the applied field is reduced to 0 mT (see Fig.~7 (b)), the $R_L^+$ domains would be present near the sample edge and $R_H^+$ domains are likely to lie in the interior. No flux jumps happen in the first two quadrants, as the domains with anti-vortices are not present. In the third quadrant, as the field gets cycled to negative values, the first flux jump happens near about -50~mT. The field profiles in Fig.~7 (c) imply that in such a circumstance, the $R_L^-$ domains would lie near the edge, while $R_L^+$ and (supercooled/metastable) $R_H^+$ domains could exist in the interior of the sample. A change in the symmetry from $R_H^+ \rightarrow R_L^+$ in some domains could trigger an additional perturbation in the dynamically varying distribution of $R_L^-$, $R_L^+$ and $R_H^+$ domains, such that the vortices and anti-vortices lying in juxtaposition in $R_L^+$/$R_L^-$ domains could start annihilating each other leading to an avalanche resulting in a flux jump.

\begin{figure}[!htb]
\begin{center}
\includegraphics[scale=0.5,angle=0] {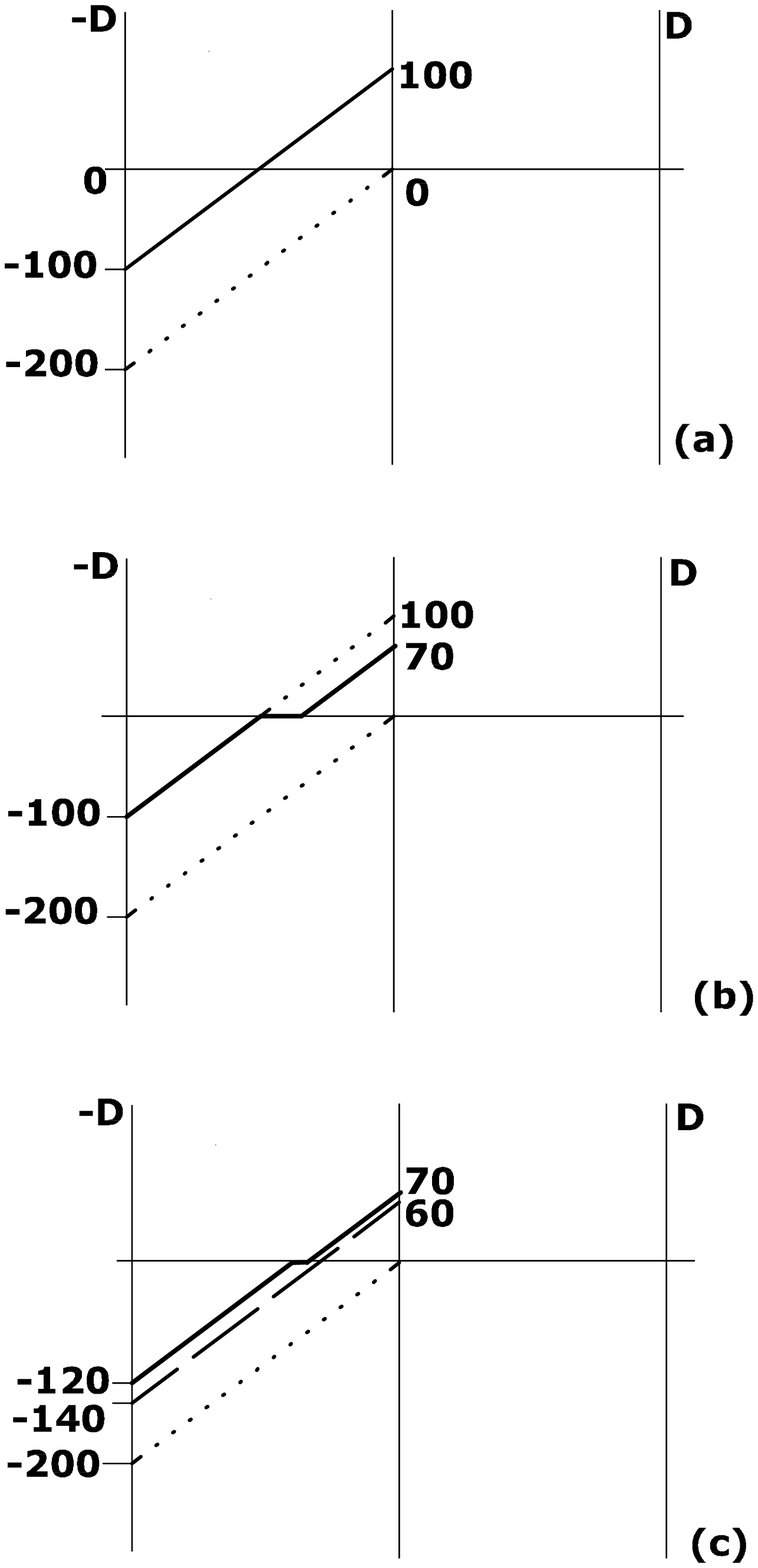}
\caption{ Bean's profiles (only one half section (-D to 0) is shown) at an applied field of $\mu_0H$ = -100 mT corresponding to a situation (a) just before the flux jump and (b) just after the flux jump. (c) Bean's profiles as the field ramps upto $\mu_0$H = -140 mT}
\label{FIG. 8.}
\end{center}
\end{figure}

From Fig.~5, one can find that the magnetization values start to build up once again after the flux jump happens, and one can envisage the notion of an underlying envelope M-H loop. As the applied field ramps up beyond -100~mT, another (more) flux jump(s) happen (cf. Fig.~5 (b)). Such jumps could possibly be triggered by $R_L^- \rightarrow R_H^-$ transition in the domains, which are nucleating near the edge of the sample. When the applied field is in the range of -100~mT to -150~mT, the $R_L^+$ domains would be present in the middle of the sample. Their presence would permit the annihilation of vortices and anti-vortices in $R_L^+$ and $R_L^-$ domains lying in juxtaposition in the interior. The recurrence of the process of $R_L^- \rightarrow R_H^-$ transition could trigger more than one avalanche, while the field is getting ramped down to -200~ mT in the third quadrant. Similar arguments will explain the flux jumps in the fifth quadrant.\\

~~~~~~We show in Fig.~8 a plausible sequence of Bean's profile, when the flux jump is located, say at H = 100~mT. As the flux jumps occurs, the Bean's profile of Fig.~8 (a) probably transforms to the profile in Fig.~8 (b), due to movement of anti-vortices from the exterior of the sample into the portion in which anti-vortices and vortices have annihilated to create a current free region at the right side of B = 0. The profile in Fig.~8 (b) implies the rearrangement of vortices in the interior of the sample, as shown by the solid line. It is apparent that profile in Fig.~8 (b) corresponds to a lower net magnetization value as compared to that for profile in Fig.~8 (a). As the field would further ramp away from -100~mT, the critical current would gradually get set up in the gradient free region. Fig~8 (c) shows that at -140~mT, the Bean's profile has assumed a form, as if the flux jump had not occurred at -100~mT, thereby implying the return of the magnetization values to the underlying envelope hysteresis loop.

\subsubsection{Tracings of the minor hysteresis curves with different thermomagnetic histories}

To check the validity of our conjecture, we traced several minor hysteresis curves with different thermomagnetic histories, including the tracings of the complete hysteresis loops after having cooled the sample in different fields. Fig.~9 shows two representative minor hysteresis curves obtained at 2.1~K in Y1221 ($\#$ A) using the field ramp rate of 0.2 T/min in the VSM. In panel (a) of Fig.~9, after zero field cooling, the field is initially ramped upto +75~mT (filled squares), it is then reversed to -500~mT (open triangles), followed by ramping up again to +75~mT (filled circles). No flux jump is observed at any magnetic field in the third quadrant, however, a flux jump occurs at about +32~mT in the fifth quadrant.\\
Panel (b) of Fig.~9 shows that if the field is reversed to -500~mT from +174~mT, two flux jumps are observed at about -91~mT and -137~mT, respectively, in the third quadrant. It is pertinent to note that a flux jump in the range of -30~mT to -50~mT is not present in the third quadrant. On ramping up the field from -500 mT to +174 mT , one can witness three flux jumps at about 37~mT, 97~mT and 142~mT, respectively, in the fifth quadrant.

\begin{figure}[!htb]
\begin{center}
\includegraphics[scale=0.4,angle=0] {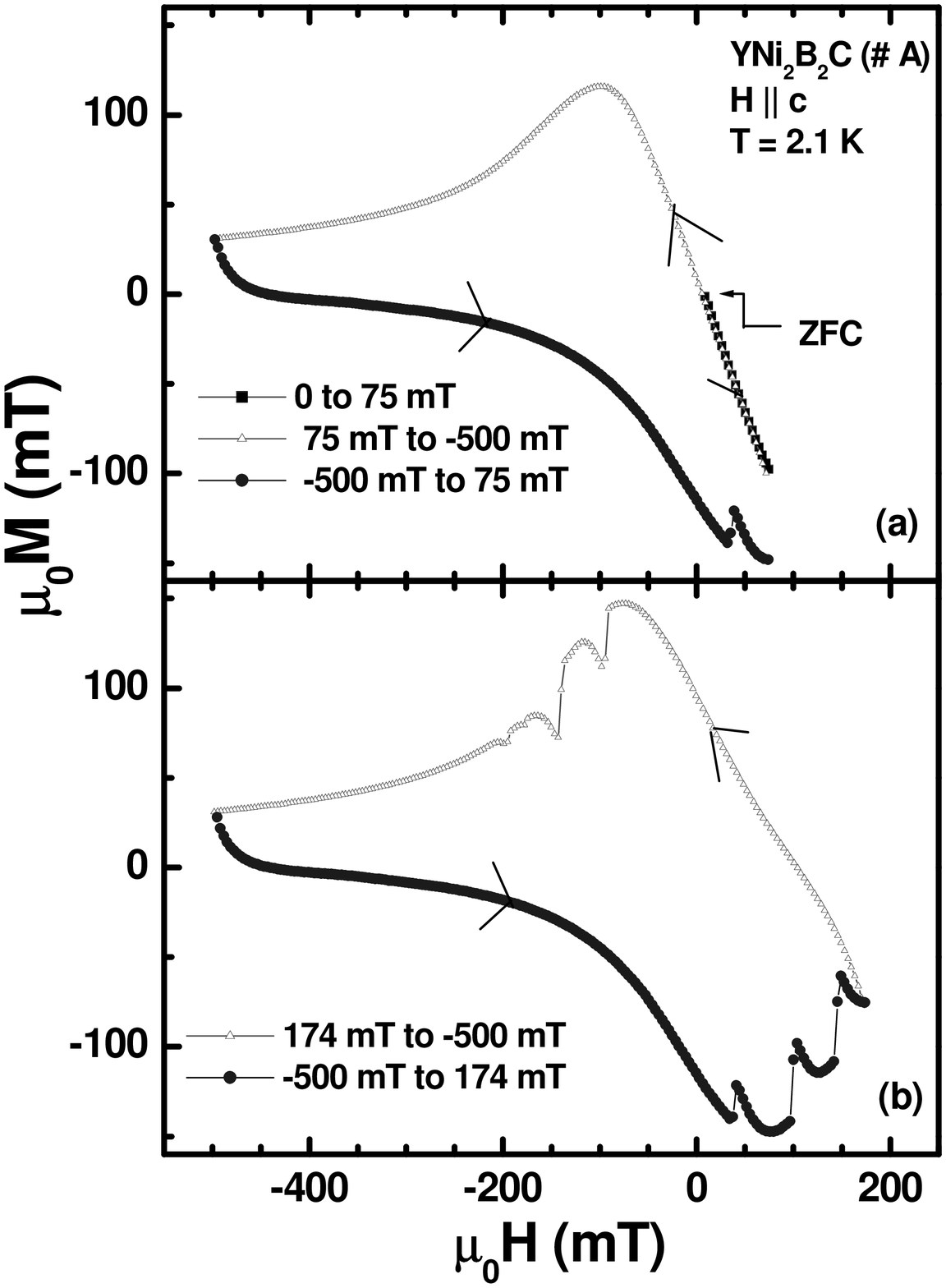}
\caption{In panel (a), the filled square data points trace the initial magnetization curve (0 to 75~mT) after zero field cooling (ZFC) at T = 2.1~K in Y1221 ($\#$ A) with H $||$ c. The open triangle and filled circle data points in different panels correspond to the minor curves traced while ramping up the field to -500~mT from (a) 75~mT and (b) 174~mT respectively, and reversing the field from -500~mT to (a) 75~mT and (b) 174~mT, respectively.}
\label{FIG. 9.}
\end{center}
\end{figure}

\begin{figure}[!htb]
\begin{center}
\includegraphics[scale=0.45,angle=90] {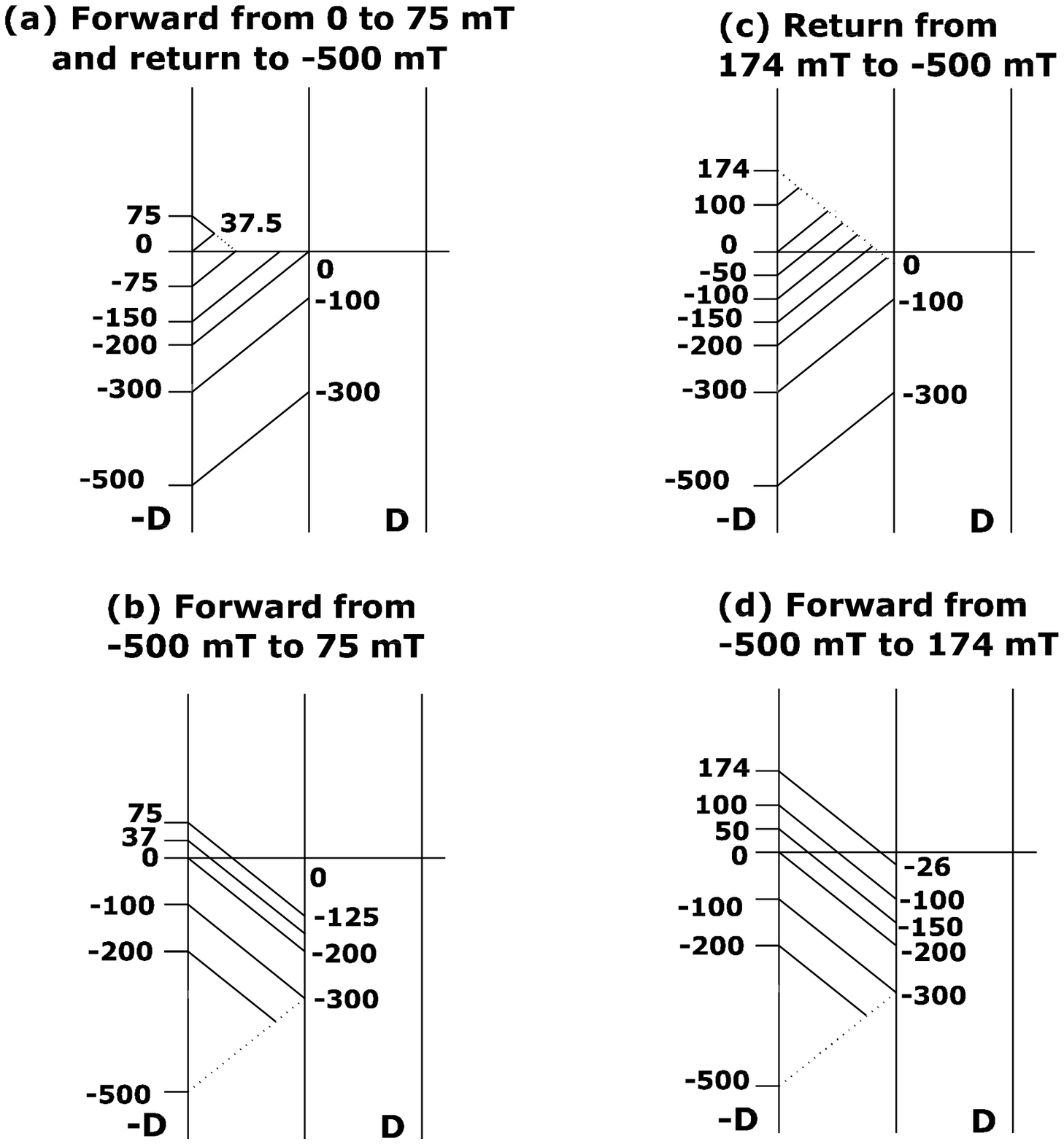}
\caption{Panels (a) and (b) correspond to Bean's profiles (only one half section (-D to 0) shown) relevant for the minor curves of Fig.~9 (a). Panels (c) and (d) show the same for the minor curves of Fig.~9 (b).}
\label{FIG. 10.}
\end{center}
\end{figure}

Fig.~10 shows the Bean's profiles corresponding to the data in Fig.~9. The profiles in Fig.~10 (a) imply that when the applied field reaches a value of -75~mT, the $R_L^-$ and $R_L^+$ domains lie in juxtaposition in the outer portion of the sample and there are no vortices of any kind in the middle region of the sample. As the applied field gets ramped from -75~mT to -150~mT, the specimen gets progressively more filled with $R_L^-$ domains. As the ramping of the field proceeds further from -150~mT to -500~mT, the $R_L^-$ to $R_H^-$ transition could (repeatedly) occur near the edge of the sample, but there are no domains with vortices of the opposite kind ($R_L^+$ like or $R_H^+$ like) in the interior. The conjecture stated above thus, precludes the occurrence of flux jump in the third quadrant in Fig.~9 (a). The profiles in Fig.~10 (b) further show that when the flux jump happens on approaching +37~mT field in Fig.~9 (a), the $R_L^+$ and $R_L^-$ domains would lie in juxtaposition near the edge of the sample and $R_H^-$ domains would lie in the deep interior, near the central region. A transition from $R_H^-$ to $R_L^-$ in the interior region probably provides the necessary trigger for annihilation of vortices across the boundaries of $R_L^-$ and $R_L^+$ domains.\\
An examination of the profiles in Fig.~10 (c) in conjunction with the flux jumps in Fig.~9 (b) would imply that as the applied field reaches a value of -50~mT in the third quadrant, the $R_L^-$ domain will lie near the edge, and only the $R_L^+$ domains would fill the interior of the sample. There is no possibility of $R_L \rightarrow R_H$ transition anywhere in the sample between 0 and -50~mT. The $R_L^- \rightarrow R_H^-$ transition probably happens near the edge of the sample as the applied field ramps towards -91~mT (see Fig.~9 (b)), at that stage, the interior of the sample contains $R_L^+$ domains. As the field ramps further to -137~mT, another $R_L^-$ to $R_H^-$ transition could occur near the edge, while the $R_L^+$ domains are still left in the interior of the sample. This could explain the two flux jumps in the third quadrant in Fig.~9 (b). The profiles in Fig.~10 (d) further show that as the applied field approaches +37~mT in the fifth quadrant, the situation is similar to that in the fifth quadrant as depicted in Fig.~10 (b) and the first flux jump in Fig.~9 (b) gets triggered at such a field, by $R_H^-$ to $R_L^-$ transition in the interior. The subsequent two jumps in the fifth quadrant in Fig.~9 (b) are triggered by the possibility of recurring $R_L^+$ to $R_H^+$ transitions near the sample edge, while $R_L^-$ domains continue to exist in the interior.\\
Field cooled (FC) measurements, in principle, result in a near uniform field distribution across the sample. In a FC(H) state, the crystal will comprise $R_L$/$R_H$ or a co-existence of $R_L$ and $R_H$ domains, depending on the field value. Tracings of minor hysteresis loops by changing the field in the third quadrant, from different $M_{FC(H)}$ values revealed (data not being shown here) one/two/three jumps. Their occurence can also be rationalized in terms of above stated conjecture by drawing appropriate Bean's field profiles.

\subsubsection{Temperature dependence of the flux jumps}

It is well documented that the tendency and the magnitude of the flux jumps decreases with an increase  in temperature \cite{rwilson45}. The plots of the M-H loops in Y1221 ($\#$ A) at higher temperatures (T $>$ 2.1~K) in Fig.~11 seem to conform to this notion. A comparison of the data in Fig.~11 with the corresponding data in Fig.~5 (a) reaffirms the premise that the extent of jump(s) decreases on progressive increase of temperature from 2.1~K to 8.45~K. A significant observation is that the flux jump evident at lower fields (50-75~mT range) in Fig.~5 (a) is not present at higher temperatures (i.e., T $\geq$ 4~K as in Fig.~11). This trend could also find a rationalization in terms of the decrease of $J_c$(H) (or full penetration field) with the increase in temperature, and the corresponding changes in the field distribution inside the sample. In Fig.~11 (b), the $H^*$ value at 6.45~K would be reckoned to be about 150~mT. Bean's profile drawn for $H^*$ = 150~mT would then demonstrate the absence of flux jumps at the lower field values.

\begin{figure}[!htb]
\begin{center}
\includegraphics[scale=0.4,angle=0] {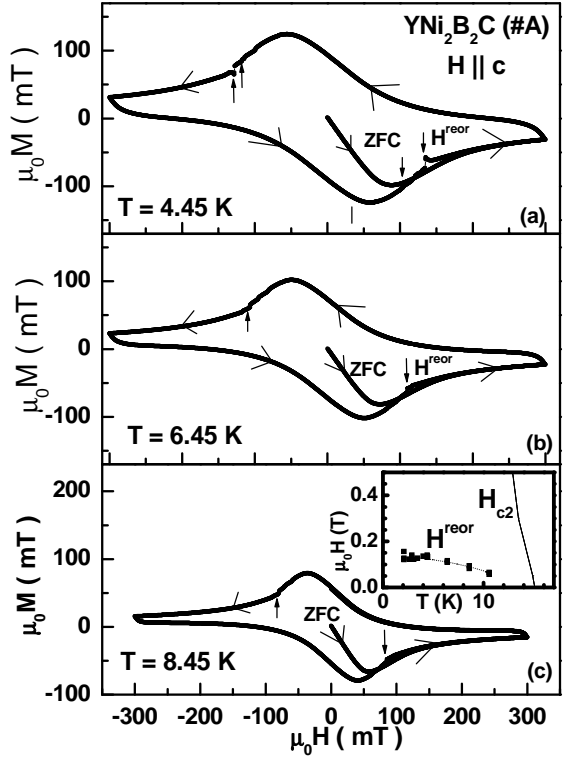}
\caption{M-H loops at 4.45~K (a), 6.45~K (b) and 8.45~K (c) in Y1221 ($\#$ A) with H $||$ c. Inset in panel (c) shows temperature variation in $H^{reor}$ and $H_{c2}$ (see text for details).}
\label{FIG. 11.}
\end{center}
\end{figure}

Our assertion that a flux jump is triggered by $R_L \rightarrow R_H$ transition and is facilitated by the juxtaposition of vortices and anti-vortices implies that the (highest) limiting field value at which a jump is observed (in a given fifth/third quadrant) has to be lower than the corresponding $H^*$ at a given temperature. Such a limit could also be taken as indicative of the field value at which $R_L \rightarrow R_H$ transition happens near the edge of the sample. Keeping this in view, we draw attention to the plot of such limiting values (designated as $H^{reor}$ and marked by arrows in panels (a) to (c) of Fig.~11) as a function of temperature in the inset of Fig.~11 (c). Multiple values of $H^{reor}$ at a given temperature represent the spread in these values during different M-H runs at the chosen temperatures. It is satisfying to see the similarity in $H^{reor}$(T) line determined as above from our flux jump data and the $H_1$(T) line (representing $R_L \rightarrow R_H$ transition) determined from SANS measurements in a crystal of Y1221 by Dewhurst {\it et al} \cite{rdewhurst27}.

\subsection{Flux jumps in $LuNi_2B_2C$}
\begin{figure}[!htb]
\begin{center}
\includegraphics[scale=0.4,angle=270] {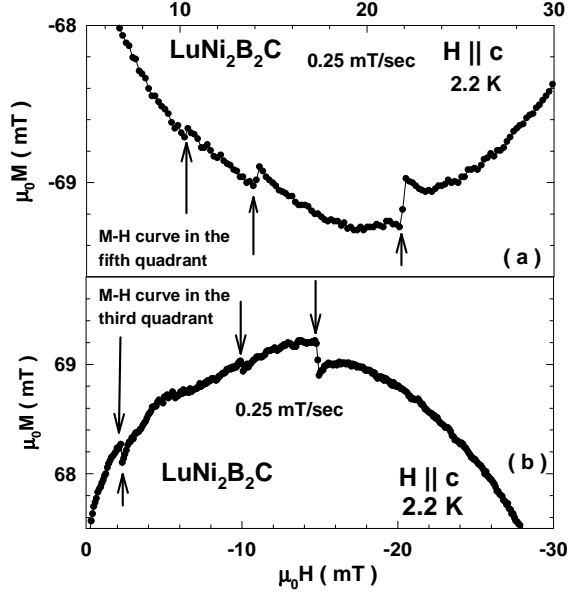}
\caption{Expanded portions of the M-H loop obtained at 2.2 K in Lu1221 with H $||$ c. Panels (a) and (b) depict the M-H curves in the fifth (ramping the field from 0 to $H_{max}$ = 200~mT) and the third (ramping the field from 0 to -$H_{max}$ = -200~mT) quadrants, respectively. The locations of the flux jumps have been marked with arrows.}
\label{FIG. 12.}
\end{center}
\end{figure}

The plots in Fig.~13 show parts of the M-H loop recorded using VSM at 2.2 K in Lu1221, for H $||$ c, with a scan rate of 0.25 mT/sec. The presence of the three flux jumps in the third and fifth quadrants in the field interval 2 to 25~mT can be noted. From Bitter decoration experiments performed in the field cooled state on single crystals of Lu1221, the $R_L \rightarrow R_H$ transition is reckoned to occur in the interval of 20 mT to 50 mT \cite{rvinnikovh125}. The observation of flux jumps in a  single crystal of Lu1221 at smaller field values as compared to those in Y1221 is encouraging. The nominal $H^*$ at 2.1~K in the given crystal of Lu1221, with H $||$ c, is about 30 mT. Simplistic reasoning based on Bean's profiles for $H^*$ = 30~mT can rationalize the occurrence of the flux jumps at the observed field values. The local macroscopic field would envisage $R_H^+$ to $R_L^+$ transition deep inside the sample and $R_L^-$ to $R_H^-$ transition near the surface as the applied field ramps from zero to -30~mT in the third quadrant.

 \subsection{Measurement of quadrupolar signal (Q) in the crystal `A' of $YNi_2B_2C$ using a VSM}

When the magnetization in a given sample is non-uniform, multipole moments other than the dipole, also, contribute \cite{rguy46} to a measured signal. Preferential measurement of the quadrupole moment (Q) is a very useful technique to gain information on the spatial inhomogeneity in the magnetization across a given sample \cite{rguy47}. Such a measurement has been performed using a VSM in Y1221 ($\#$ A), with H $||$ c (see Fig.~14) to explore the fingerprints of the notion of the flux jumps observed in the M-H loops. The details of the measurement procedure can be found elsewhere \cite{rdilip48,rme49}. In brief, a sample is moved from the central region of the astatic pair of the coils of VSM to another location, where the signal due to the dipole moment is expected to cross from a positive to a negative value. At such a location, the measured signal preferentially captures the contribution from the quadrupolar moment of the sample.

\begin{figure}[!htb]
\begin{center}
\includegraphics[scale=0.4,angle=0] {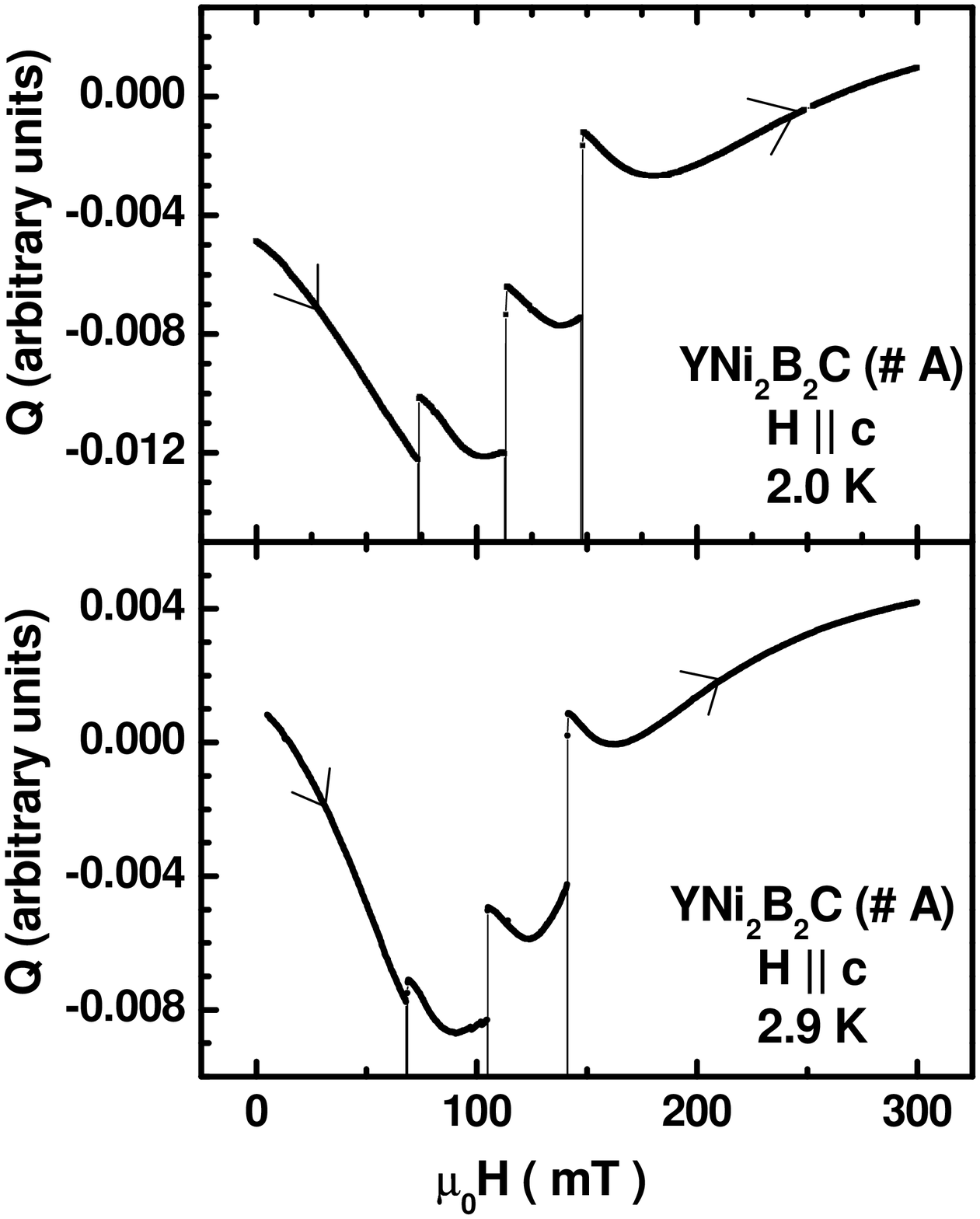}
\caption{The panels (a) and (b) depict the plots of quadrupolar signal {\bf Q} (in arbitrary units) vs. H in Y1221 ($\#$ A) for H $||$ c at T = 2 K and 2.9 K, respectively.}
\label{FIG. 13.}
\end{center}
\end{figure}
Panels (a) and (b) of Fig. 14 show Q (in arbitrary units) vs. H plots at 2.0 K and 2.9 K, respectively as the field is sweeped from zero to $H_{max}$ in the so called fifth quadrant at a sweep rate of 0.01 T/minute. The Q vs. H data in Fig. 14 (a) appears to have a shape similar to the M vs. H data in Fig. 5, and this could imply that the residual contribution from the dipole moment of the sample is still dominating the measured signal at the new preferred location for the record of the quadrupolar contribution. There is, however, one notable observation in both the panels of Fig. 14, which purports to support the significant presence of the contribution from the quadrupolar moment of the sample at the new location. It may be mentioned that at the field values of the flux jumps, the measured signal gets out of the range of the plot with the Lock-in amplifier settings getting overboard momentarily and it returns to within the range only when the process of the flux jump is complete. A motivated search for such an occurrence (i.e., signal getting out of range) in the magnetization response at the usual central position of the astatic pair of the coils, did not yield an affirmative answer. This implies that the residual presence of the dipolar signal at the preferred position for the Q measurement, is not responsible for the signal getting out of range, when the flux jump happens.\\
We are tempted to conjecture that a readjustment in the vortex matter during the flux jump process causes the contribution from the quadrupolar moment to undergo a peak like behavior. We believe that a study of anomalous variations in the quadrupolar signal, conveniently measurable in a VSM, has the potential to reveal the changes in the state of the vortex matter, which may not get fingerprinted as anomalous variations in the magnetization hysteresis response, like, the SMP anomaly and/or the peak effect phenomenon.

\subsection{Phase Diagrams in $YNi_2B_2C$ and $LuNi_2B_2C$ for H $||$ c}

Collating all the data together, we draw the phase diagram for Y1221 ($\#$ A) and Lu1221 for H $||$ c. The diagram comprises $H^{reor}$, $H_{smp}^{on}$, $H_p^{on}$ and $H_{c2}$ lines. The $H_{smp}^{on}$(T) lines have been drawn over a limited temperature interval over which data pertaining to it are presently available in these crystals. It can be noted that $H_{smp}^{on}$ line for Y1221 is flat, while it decreases with temperature for Lu1221. The former behaviour is reminiscent of the SMP line in $Ca_3Rh_4Sn_{13}$  \cite{rshampa16,rshampa17} and in optimally doped $Bi_2Sr_2CaCu_2O_8$ crystals \cite{rkhaykovich4}, while the latter is similar to the SMP line in 2H-$NbSe_2$ crystal \cite{rAjay50}.

\begin{figure}[!htb]
\begin{center}
\includegraphics[scale=0.4,angle=0] {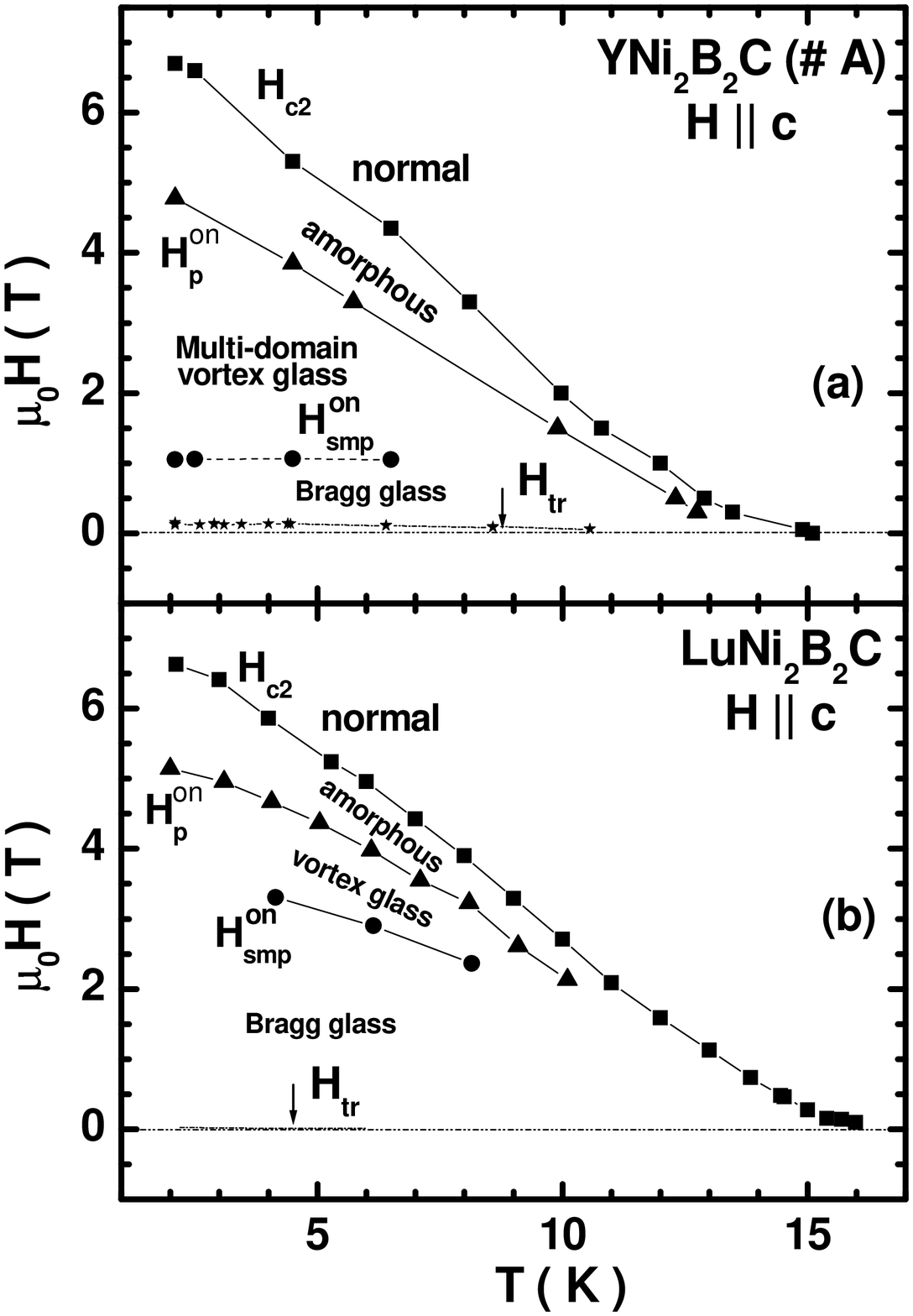}
\caption{Vortex phase diagrams in $YNi_2B_2C$ ($\#$ A) and $LuNi_2B_2C$with H $||$ c. $H^{reor}$, $H_{smp}^{on}$, $H_p^{on}$ and $H_{c2}$ lines have been sketched and different phases have been named..}
\label{FIG. 14.}
\end{center}
\end{figure}

The SMP anomaly is believed to mark a transition from a dislocation free elastic BG phase to a multi-domain VG phase  \cite{rledussal3,rgingras18}. The region below $H_{smp}^{on}$ and above $H^{reor}$ marks the BG phase. From Fig~15, it is also evident that the BG state spans over a larger (H,T) space in Lu1221 as compared to that in Y1221, attesting to the fact that lesser residual disorder prevails in the former sample.

\section{SUMMARY}

To summarize, we have expounded the details of magnetization hysteresis loops in single crystals of $YNi_2B_2C$ for H $||$ c. In the high field regime, a second magnetization peak anomaly and the peak effect are observed, while at lower fields (H $<$ 200~mT), the flux jumps are evident. The flux jumps have been noted in minor hysteresis curves recorded with differing thermomagnetic histories, in order to explore a probable reason for their occurrence. On the basis of Bean's profiles sketched for different thermomagnetic histories, we have conjectured that the flux jumps get triggered as a result of the structural transitions, $R_{L,H} \rightarrow R_{H,L}$, possible in the vortex lattice in the field range 100-150~mT for H $||$ c in $YNi_2B_2C$. Similar flux jumps have also been observed in a cleaner crystal (as compared to those of $YNi_2B_2C$) of $LuNi_2B_2C$ at lower fields (H $<$ 25~mT), giving support to an assertion made by us. Another evidence in support of our conjecture could come from the measurement of quadrupole signal (Q) in $YNi_2B_2C$ using a VSM. Q undergoes a peak like feature at applied field values, corresponding to which the macroscopic field somewhere inside the sample is such that $R_{L,H} \rightarrow R_{H,L}$ transition happens. The onset of a SMP anomaly at a lower field in the crystal of $YNi_2B_2C$ as compared to that in $LuNi_2B_2C$ could be termed as consistent with the occurrence of  $R_L \rightarrow R_H$ transition at a higher field in it. Based on all the measurements, vortex phase diagrams for Y1221 ($\#$ A) and Lu1221 for H $||$ c have been drawn, which depict various phases of vortex matter. Our results call for a study to map Bean's profiles by local micro-Hall bar arrays \cite{rjames41} in all the five quadrants for different thermomagnetic histories, in samples which display flux jumps.

\section*{ACKNOWLEDGMENTS}
 Three of us (DJ-N, ADT and DP) would like to thank the TIFR Endowment Fund for the Kanwal Rekhi Career Development support. We would like to thank Morten Eskildsen (University of Notre Dame, U.S.A.) for many useful inputs and valuable discussions, in particular, his clarifying account of the changes in field profiles, while a flux jump happens. The $LuNi_2B_2C$ crystals were obtained from Paul Canfield at Ames Laboratory, Iowa State University, Ames, U.S.A. We also gratefully acknwledge Rinke Wijngaarden (Vrije Universiteit, Amsterdam, The Netherlands) for kindly sharing their observations on flux jumps in the samples of Pb.


\begin{references}

\bibitem{rblatter1} G. Blatter, M. V. Fiegel'man, V. B. Geshkenbien, A. I. Larkin, and V. M. Vinokur, Rev. Mod. Phys. {\bf 66}, 1125 (1994) and the references therein.

\bibitem{rgiamarchi2} T. Giamarchi and S. Bhattacharya, in High Magnetic Fields: Applications to Condensed Matter Physics and Spectroscopy, C. Bertier, L.P. Levy and G. Martinez, eds., p314 (Springer Verlag, 2002) and the references therein.

\bibitem{rledussal3} T. Giamarchi, and P. Le Doussal, Phys. Rev. Lett {\bf 72}, 1530 (1994); Phys. Rev. B {\bf 52}, 1242 (1995).

\bibitem{rkhaykovich4} B. Khaykovich, E. Zeldov, D. Majer, T. W. Li, P. H. Kes, and M. Konczykowski, Phys. Rev. Lett. {\bf 76}, 2555 (1996).

\bibitem{rsatyajit5} S. S. Banerjee, A. K. Grover, M. J. Higgins, Gautam I. Menon, P. K. Mishra, D. Pal, S. Ramakrishnan, T. V. Chandrasekhar Rao, G. Ravikumar, V. C. Sahni, S. Sarkar, and C.V. Tomy, Physica C {\bf 355}, 39 (2001) and the references therein.

\bibitem{rcheon6} K. O. Cheon, I. R. Fisher, V. G. Kogan, P. C. Canfield, P. Miranovic, and P. L. Gammel, Phys. Rev. B {\bf 58}, 6463 (1998).

\bibitem{rklein7} T. Klein, I. Joumard, S. Blanchard, J. Marcus, R. Cubitt, T. Giamarchi, and P. Le Doussal, Nature(London) {\bf 413}, 404 (2001).

\bibitem{rdivakar8} U. Divakar, A. J. Drew, S. L. Lee, R. Gilardi, J. Mesot, F. Y. Ogrin, D. Charalambous, E. M. Forgan, G. I. Menon, N. Momono, M. Oda, C. D. Dewhurst, and C. Baines, Phys. Rev. Lett. {\bf 92}, 237004 (2004).

\bibitem{rsoibel9} A. Soibel, E. Zeldov, M. Rappaport, Y. Myasoedov, T. Tamegai, S. Ooi, M. Konczykowski, and V. B. Geshkenbein, Nature(London) {\bf 406}, 282 (2000).

\bibitem{ravraham10} N. Avraham, B. Khaykovich, Y. Myasoedov, M. Rappaport, H. Shtrikman, D. E. Feldman, T. Tamegai, P. H. Kes, M. Li, M. Konczykowski, K. van der Beek, and E. Zeldov, Nature(London) {\bf 411}, 451 (2001).

\bibitem{rdilip11} D. Pal, S. Ramakrishnan, A. K. Grover, D. Dasgupta, and B. K. Sarma, Phys. Rev. B {\bf 63}, 132505 (2000).

\bibitem{rdilip12} D. Pal, S. Ramakrishnan, and A. K. Grover, Phys. Rev. B  {\bf 65}, 096502 (2001).

\bibitem{rling13} X. S. Ling, S. R. Park, B. A. McClain, S. M. Choi, D. C. Dender, and J.W. Lynn, Phys. Rev. Lett. {\bf 86}, 712 (2001).

\bibitem{rpark14} S. R. Park, S.M. Choi, D. C. Dender, J.W. Lynn, and X. S. Ling, Phys. Rev. Lett. {\bf 91}, 167003 (2003).

\bibitem{troyanovski} A. M. Troyanovski, M. van Hecke, N. Saha, J. Aarts, and P. H. Kes, Phys. Rev. Lett. {\bf 89}, 147006 (2002).

\bibitem{rdaeumling15} M. Daeumling, J. M. Seuntjens, and D. C. Larbalestier, Nature(London) {\bf 346}, 332 (1990).

\bibitem{rshampa16} S. Sarkar, P. L. Paulose, S. Ramakrishnan, A. K. Grover, C. V. Tomy, G. Balakrishnan, and D. McK. Paul, Physica C {\bf 356}, 181 (2001).

\bibitem{rshampa17} S. Sarkar, D. Pal, P. L. Paulose, S. Ramakrishnan, A. K. Grover, C. V. Tomy, D. Dasgupta, Bimal K. Sarma, G. Balakrishnan, and D. McK. Paul, Phys. Rev. B  {\bf 64}, 144510 (2001).

\bibitem{rgingras18} M. J. P. Gingras and D. A. Huse, Phys. Rev. B  {\bf 53}, 15193 (1996).

 \bibitem{rabrikosov19} A.A. Abrikosov, Sov. Phys.- JETP {\bf 5}, 1174 (1957).

\bibitem{robst20} B. Obst, Phys. Status Solidi B {\bf 45}, 467 (1971) and the references therein.

\bibitem{rcanfield21} P.C. Canfield, P. L. Gammel and D. J. Bishop, Phys. Today {\bf 51}(10), 40 (1998) and the references therein.

\bibitem{reskildsen22} M. R. Eskildsen, P. L. Gammel, B. P. Barber, A. P. Ramirez, D. J. Bishop, N. H. Andersen, K. Mortensen, C. A. Bolle, C. M. Lieber, and P. C. Canfield, Phys. Rev. Lett. {\bf 79}, 487 (1997).

\bibitem{rmcpaul23} D. Mck. Paul, C.V. Tomy, C.M. Aegerter, R. Cubitt, S.H. Lloyd, E.M. Forgan, S.L. Lee, and M. Yethiraj, Phys. Rev. Lett. {\bf 80}, 1517 (1998).

\bibitem{rvinnikovh224} L. Ya. Vinnikov, T.L. Barkov, P.C. Canfield, S.L. Bud'ko, and V.G. Kogan, Phys. Rev. B {\bf 64}, 024504 (2001).

\bibitem{rvinnikovh125} L. Ya. Vinnikov, T.L. Barkov, P.C. Canfield, S.L. Bud'ko, J. E. Ostenson, F. D. Laabs, and V. G. Kogan, Phys. Rev. B {\bf 64},  220508 (2001).

\bibitem{rlevett26} S. J. Levett, C. D. Dewhurst, and D. McK. Paul, Pramana-Journal of Phys. {\bf 58}, 913 (2002).

\bibitem{rdewhurst27} C. D. Dewhurst, S. J. Levett, and D. McK. Paul, Phys. Rev. B {\bf 72}, 014542 (2005).

\bibitem{rnag28} R. Nagarajan, Chandan Mazumdar, Zakir Hossain, S. K. Dhar, K. V. Gopalakrishnan, L. C. Gupta, C. Godart, B. D. Padalia, and R. Vijayaraghavan, Phys. Rev. Lett. {\bf 72}, 274 (1994).

\bibitem{rcava29} R. J. Cava, H. Takagi, H. W. Zandbergen, J. J. Krajewski, W. F. Peck, T. Siegrist, B. Batlogg, R. B. van Dover, R. J. Felder, K. Mizuhashi, J. O. Lee, H. Eisaki, and S. Uchida, Nature (London) {\bf 367}, 252 (1994).

\bibitem{ryethiraj30} M. Yethiraj, D. K. Christen, D. McK. Paul, P. Miranovic, and J. R. Thompson, Phys. Rev. Lett. {\bf 82}, 5112 (1999).

\bibitem{rgilardi31} R. Gilardi, J. Mesot, A. Drew, U. Divakar, S. L. Lee, E. M. Forgan, O. Zaharko, K. Conder, V. K. Aswal, C. D. Dewhurst, R. Cubitt, N. Momono, and M. Oda, Phys. Rev. Lett. {\bf 88}, 217003 (2002).

\bibitem{rrosenstein32} B. Rosenstein, B. Ya. Shapiro, I. Shapiro, Y. Bruckental, A. Shaulov, and Y. Yeshurun, Phys. Rev. B  {\bf 72}, 144512 (2005).

\bibitem{rforgan33} M. Laver, E. M. Forgan, S. P. Brown, D. Charalambous, D. Fort, C. Bowell, S. Ramos, R. J. Lycett, D. K. Christen, J. Kohlbrecher, C. D. Dewhurst and R. Cubitt hys. Rev. Lett. {\bf 96}, 167002 (2006).

\bibitem{rbrown34} S. P. Brown, D. Charalambous, E. C. Jones, E. M. Forgan, P. G. Kealey, A. Erb, and J. Kohlbrecher, Phys. Rev. Lett. {\bf 92}, 067004 (2004).

\bibitem{eskildsen} M. R. Eskildsen, A. B. Abrahamsen, D. Lpez, P. L. Gammel, D. J. Bishop, N. H. Andersen, K. Mortensen, and P. C. Canfield, Phys. Rev. Lett. {\bf 86}, 320 (2000).

\bibitem{eskildsenpra} M. R. Eskildsen, A. B. Abrahamsen, P. L. Gammel, D. J. Bishop, N. H. Andersen, K. Mortensen, and P. C. Canfield, Pramana-J. Phys. {\bf 58}, 903 (2002).

\bibitem{rkogan35} V. G. Kogan, M. Bullock, B. Harmon, P. Miranovic, Lj. Dobrosavljevic-Grujic, P. L. Gammel and D. J. Bishop, Phys. Rev. B  {\bf 55}, R8693 (1997); V. G. Kogan, P. Miranovic', Lj. Dobrosavljevic-Grujic, W. E. Pickett, and D. K. Christen, Phys. Rev. Lett. {\bf 79}, 741 (1997).

\bibitem{rgammel36} P. L. Gammel, D. J. Bishop, M. R. Eskildsen, K. Mortensen, N. H. Andsersen, I. R. Fisher, K. O. Cheon, P. C. Canfield, and V. G. Kogan, Phys. Rev. Lett. {\bf 82}, 4082 (1999).

\bibitem{rsilhane37} A. V. Silhanek, J. R. Thompson, L. Civale, D. McK. Paul, and C. V. Tomy, Phys. Rev. B {\bf 64}, 012512 (2001).

\bibitem{rxu38} Ming Xu, P. C. Canfield, J. E. Ostenson, D. K. Finnemore, B. K. Cho, Z. R. Wang and D. C. Johnston, {\it Physica}  {\bf C227}, 321 (1994).

\bibitem{rhirata39} K. Hirata, H. Takeya, T. Mochiku and K. Kadowaki in: {\it Proceedings of the $8^{th}$ International Symposium on Superconductivity}, edited by H. Hayakawa and Y. Enomoto (Springer-Verlag, Tokyo, 1996), p. 619.

\bibitem{rsong40} K. J. Song, J. R. Thompson, M. Yethiraj, D. K. Christen, C. V. Tomy and D. McK. Paul,   Phys. Rev. B {\bf 59}, R6620 (1999).

\bibitem{rjames41} S. S. James, C. D. Dewhurst, R. A. Doyle, D. McK. Paul, Y. Paltiel, E. Zeldov and A. M. Campbell, Physica C  {\bf 332}, 173 (2000).

\bibitem{rtakeya42} H. Takeya, T. Hirano and K. Kadowaki, Physica C {\bf 256}, 220 (1996).

\bibitem{rinke} R. Wijngaarden et al., unpublished results.

\bibitem{rbean43} C. P. Bean, Rev. Mod. Phys. {\bf 36}, 31 (1964).

\bibitem{rchaddah44} P. Chaddah, K. V. Bhagwat, and G. Ravikumar, Physica C {\bf 159}, 570 (1989); P. Chaddah, in {\it Studies of High Temperature Superconductors}, edited by A. V. Narlikar, vol. {\bf 14}, (Nova Science Publishers, New York, U.S.A., 1995), p. 245

\bibitem{rwilson45} M. N. Wilson, in {\it Superconducting Magnets}, Clarendon Press Oxford (1983).

\bibitem{rguy46} C. N. Guy and W. Howarth, J. Phys. C: Solid State Phys. {\bf 11}, 1635 (1978).

\bibitem{rguy47} C.N. Guy, J. Phys. F {\bf 12}, 1453 (1982).

\bibitem{rdilip48} D. Pal, S. Ramakrishnan, A. K. Grover, M. J. Higgins and M. Chandran, Physica C {\bf 369}, 200 (2002).

\bibitem{rme49} D. Jaiswal-Nagar, A. D. Thakur, M. R. Eskildsen, P. C. Canfield, S. M. Yusuf, S. Ramakrishnan and A. K. Grover, Physica B  {\bf 359-361}, 476 (2005).

\bibitem{rAjay50} A. D. Thakur,  S. S. Banerjee, M.J. Higgins, S. Ramakrishnan, and A. K. Grover, Phys. Rev. B  {\bf 72}, 134524 (2005).

\end{references}
\end{document}